\documentclass{article}
\usepackage{amsfonts}
\usepackage{amsmath}

\begin{document}
\newtheorem{veta}{Theorem}
\newtheorem{lema}{Lemma}
\newtheorem{definice}{Definition}

\title{On the Hadamard condition on Robertson-Walker spacetime}

\author{Jan Tichavsk\'y}

\maketitle

\begin{abstract}
Using construction of adiabatic vacuum states of a free scalar field 
on Robertson-Walker spacetime, using results of L\"uders and Roberts
 we prove that validity of the Hadamard condition implies smoothness
of the scale factor.
\end{abstract}

\section{Introduction} 

In the end of the sixties Parker \cite{pa1}, \cite{pa2} investigated the particle creation in 
dynamic universe using adiabatic vacuum states. Later,  L\"uders and Roberts gave an exact mahematical
definition in \cite{lr} based on the algebraic approach to QFT of Haag, Kastler \cite{hk}.
In their article the Klein-Gordon equation is solved by the Fourier method. As a part of this, they found 
the equation for the time dependent part of the state distribution, which lead to the another 
equation whose solutions describes adiabatic vacuum states, and as a limit case the Hadamard states.
They don't solve this equation exactly, but by iterations. All iterative solutions were obtained
under the assumption of smoothness of the scale factor.\\
In the first part of this paper we give a construction of adiabatic vacuum states and Hadamard states
according to L\"uders and Roberts \cite{lr}. In the second part we will give a statement 
of the Hadamard condition, in the way of characterization
of the microlocal structure $2$-point Wightman distribution due to Radzikowski \cite{rad}.
Here we assume the validity of the Hadamard condition and its corollaries to prove that
the scale factor is smooth.

\section{Hadamard Condition for Linear Klein-Gordon Field}

Let $M$ be a globally hyperbolic four dimensional Riemannian manifold 
with metric tensor $g$ and Riemannian connection $\nabla$. Consider a
 scalar field $\Phi:M\to \mathbb {R}$ satisfying the Klein-Gordon equation
\begin{equation}
\label{kg}
(\Box_{g}+m^{2})\Phi=0.
\end{equation}
The hyperbolicity of $M$ implies that this equation is well-posed. Denote by $C^{\infty}_{0}(M)$
the space of smooth functions with compact support  on $M$ and by $C^{\infty}(M)$ the space
of smooth functions on $M$. There are two uniquely determined continous operators
$$\Delta_{R},\Delta_{A}:C^{\infty}_{0}(M)\to C^{\infty}(M),$$
such that
$$(\Box_{g}+m^{2})\Delta_{R}f = \Delta_{R}(\Box_{g}+m^{2})f=f,$$
and similarly for $\Delta_{A}$, and
$$\begin{array}{rcl}
{\rm supp}\ (\Delta_{A}f)&\subset & J^{-}({\rm supp}\ f),\\
{\rm supp}\ (\Delta_{R}f)&\subset & J^{+}({\rm supp}\ f),
\end{array}$$
for $f\in C^{\infty}_{0}(M)$, where $J^{+}(S)$ means the causal future and $J^{-}(S)$ the causal past of a set 
$S\subset M$. These operators are called the retarded ($\Delta_{R}$) and
the advanced ($\Delta_{A}$) propagator. Their difference $E=\Delta_{R}-\Delta_{A}$
is the propagator of the Klein-Gordon equation. The following is valid
$$\begin{array}{rcl}
(\Box_{g}+m^{2})Ef & = & E(\Box_{g}+m^{2})f=0,\\
{\rm supp}\ (Ef)&\subset & J^{+}({\rm supp}\ f)\cup J^{-}({\rm supp}\ f),
\end{array}$$
for $f$ as above.
The operators $\Delta_{R}$, $\Delta_{A}$ and $E$ have continous extensions to operators
 $\Delta_{R}',\Delta_{A}',
E':\mathcal {E'}(M)\to\mathcal {D'}(M)$, where $\mathcal {D'}(M)$ resp. 
$\mathcal {E'}(M)$ is the space of the distributions resp. the space of distributions  with compact
support.\\
Now let $\Sigma$ be an arbitrary Cauchy hypersurface in $M$ with unit normal field $n^{\alpha}$
directed to the future cone. Then there exist operators
$$\begin{array}{rclc}
\rho_{0}: & C^{\infty}(M) & \to &  C^{\infty}(\Sigma)\\
\  & f & \mapsto & f|_{\Sigma}\\
\rho_{1}: & C^{\infty}(M) & \to &  C^{\infty}(\Sigma)\\
\  & f & \mapsto & (n^{\alpha}\nabla_{\alpha}f)|_{\Sigma},
\end{array}$$
with adjoints $\rho_{0}',\rho_{1}':\mathcal {E'}(\Sigma)\to\mathcal {E'}(M)$, 
see (e. g.\cite{jun},\cite{d}).\\
Let us note that using these operators we can construct, if there are given Cauchy initial
conditions $u_{0}$, $u_{1}\in C^{\infty}_{0}(\Sigma)$, solutions of Klein-Gordon
equation (for the details se \cite{d}).\\
This makes it to possible describe elements of the phase space using initial data on the Cauchy surface
$\Sigma$. Suppose that $d^{3}\sigma$ is the volume form on this surface and define the real
symplectic space $(\Gamma,\varsigma)$, where $\Gamma = C^{\infty}(\Sigma)\otimes
C^{\infty}(\Sigma)$ is formed by the initial data of (\ref{kg}) and the real valued 
symplectic form $\varsigma$ is given by
$$\varsigma\left(\left(^{u_{1}}_{p_{1}}\right),\left(^{u_{2}}_{p_{2}}\right)\right)
=-\int_{\Sigma}[u_{1}p_{2}-u_{2}p_{1}]\ {\rm d}^{3}\sigma.$$
To this symplectic space $(\Gamma, \varsigma)$ there exists an associated Weyl algebra
$\mathcal {A}[\Gamma,\varsigma]$, generated by elements $W(F)$, $F\in\Gamma$, subject to the
relations
$$W(F)^{*}=W(F)^{-1}=W(-F),$$
$$W(F_{1})W(F_{2})=e^{-\frac{i}{2}\sigma (F_{1}, F_{2})}W(F_{1}+F_{2}),
\ \mbox {for all}\  F_{1},F_{2}\in\Gamma.$$
This Weyl algebra is a local algebra of observables in the sense of Haag and Kastler \cite{hk}.
States  on the algebra $\mathcal {A}$ are defined as complex-valued functionals on 
$\mathcal {A}$.\\
Now we can define the {\em quasifree state} $\omega_{\mu}$ associated with $\mu$ by
$$\omega_{\mu}(W(F))=e^{-\frac{1}{2}\mu (F, F)},$$
where $\mu$ is a real-valued scalar product on $\Gamma$ satisfiyng
$$\frac{1}{4}|\varsigma (F_{1}, F_{2})|^{2}\leq \mu (F_{1}, F_{1})\mu (F_{2}, F_{2}).$$
So we are now in a position to introduce two point function
$$\lambda^{(2)}(F_{1},F_{2})=\mu (F_{1}, F_{2})+\frac{i}{2}\varsigma (F_{1}, F_{2}),$$
and the Wightman the two-point distribution
\begin{equation}
\label{2b}
\Lambda^{(2)}(f_{1}, f_{2})=\lambda^{(2)}\left(\left(^{\rho_{0}Ef_{1}}_{\rho_{1}Ef_{1}}\right)
,\left(^{\rho_{0}Ef_{2}}_{\rho_{1}Ef_{2}}\right)\right),
\end{equation}
which enters into the definition the Hadamard state.\\
To introduce the latter, we use the microlocal formulation discovered by Radzikowski in \cite{rad}.
(An overview of the theory of pseudodifferential operators, microlocal analysis and wavefront sets
can be found in the book \cite{gs}.)
Denote by $T^{*}M$ the cotangent bundle of $M$. For $(x_{i},\xi_{i})
\in T^{*}M$, $i=1,2$, writing $(x_{1},\xi_{1})\sim (x_{2},\xi_{2})$ means that there is
a null geodesic $\gamma$ such that $x_{1},x_{2}\in\gamma$, with $\xi_{1}^{\mu}$ tangent
to $\gamma$ at $x_{1}$ and $\xi_{2}^{\mu}$ is obtained from $\xi_{1}$ by a parallel transport to $x_{2}$ 
along $\gamma$.
We use Theorem 3.9 of \cite{jun} as the microlocal formulation of a the Hadamard condition.

\begin{definice}
A quasifree state of a Klein-Gordon field on a  globally hyperbolic spacetime is a
{\em Hadamard state} if and only if the wavefront set of the two-point distribution (\ref{2b})
has the form
$$WF(\Lambda^{2})=\{(x_{1},\xi_{1};x_{2},\xi_{2})\in T^{*}(M\times M)\setminus
\{0\};(x_{1},\xi_{1})\sim (x_{2},\xi_{2}),\xi_{1}^{0}\geq 0\}.$$
\end{definice}
The same microlocal structure has $2$-point distribution of the Hadamard state.
Note that the Hadamard state is defined locally, but according to the "local-to-global singularity
theorem" the local Hadamard condition implies the global Hadamard condition \cite{rad1}.

\section{Adiabatic Vacuum and Hadamard States on Robetson-Walker Spacetimes}

Let $M=\mathbb {R}\times\Sigma$ be a Lorentz manifold equipped with the Robertson-
Walker metric
\begin{equation}
\label{metr}
ds^{2}=dt^{2}-a^{2}(t)\left[\frac{dr^{2}}{1-\kappa r^{2}}+
r^{2}(d\theta^{2}+\sin^{2}\theta d\varphi^{2})\right],
\end{equation}
where $\varphi\in[0, 2\pi]$, $\theta\in [0, \pi]$, $r\in [0, \infty)$ 
if $\kappa=-1, 0$ and $\varphi\in[0, 2\pi]$, $\theta\in [0, \pi]$, $r\in [0, 1)$ if
$\kappa=1$ with $a(t)\in C^{2}(\mathbb {R})$,  $a(t)>0$ for all $t\in\mathbb{R}$.
The function $a(t)$ is called a {\em scale factor}.
The Riemannian manifolds $\Sigma^{\kappa}$ are defined as
 $$\begin{array}{rcl}
 \Sigma^{+}&=&\left\{x\in\mathbb {R}^{4};\ (x^{0})^{2}+\sum_{i=0}^{3}(x^{i})^{2}=1\right\},\\
 \Sigma^{0}&=&\left\{x\in\mathbb {R}^{4};\ (x^{0})=0\right\},\\
 \Sigma^{-}&=&\left\{x\in\mathbb {R}^{4};\ (x^{0})^{2}-\sum_{i=0}^{3}(x^{i})^{2}=1, x^{0}>0\right\}.
 \end{array}$$
 On $\Sigma^{\kappa}$ we consider the metric tensor
 $$s_{ij}=\left(\begin{array}{ccc}
 \frac{1}{1-\kappa r^{2}} & \        & \ \\
 \                                   & r^{2}  & \ \\
  \                                   & \       & r^{2}\sin^{2}\theta 
 \end{array}\right).$$
Cauchy surfaces are of the form $\Sigma_{t}=\{t\}
\times\Sigma$ with $n^{\alpha}=(1,0,0,0)$. The hypersurfaces $\Sigma_{t}=\{t\}\times\Sigma$ are 
therefore equipped with the metrics $h_{ij}=a^{2}(t)s_{ij}$ in which 
we will study the Klein-Gordon equation on the $\Sigma_{t}$.
$$(\Box_{g}+m^{2})\Phi=\frac{\partial^{2}\Phi}{\partial t^{2}}+3\frac{\dot a(t)}{a(t)}
\frac{\partial\Phi}{\partial t}+(-^{(3)}\Delta_{h}+m^{2})\Phi=0,$$
where $^{(3)}\Delta_{h}$ is the Laplace-Beltrami operator on the Cauchy surface $\Sigma$,
$$^{(3)}\Delta_{h}=\frac{1}{a^{2}(t)}\left\{(1-r^{2})\frac{\partial^{2}}{\partial r^{2}}
+\frac{2-3r^{2}}{r}\frac{\partial}{\partial r}+\frac{1}{r^{2}}\Delta (\theta, \varphi)
\right\}$$
$$\Delta (\theta, \varphi)=\frac{1}{\sin\theta}\left[\frac{\partial}{\partial\theta}
\left(\sin\theta\frac{\partial}{\partial\theta}\right)+\frac{1}{\sin\theta}
\frac{\partial^{2}}{\partial\varphi^{2}}\right].$$
This is a linear partial differential equation solvable by the Fourier method. By separation
of variables we get
$$\Phi(t,r,\theta,\varphi)=\int T_{k}(t)\phi_{k}(r,\theta,\varphi)d\mu(k),$$
where $T_{k}(t)$ is obtained as the solution of the ordinary differential equations
\begin{equation}
\label{te}
\ddot T_{\vec k}(t)+3\frac{\dot a(t)}{a(t)}\dot T_{\vec k}(t)+\omega^{2}_{k}(t)T_{\vec k}(t)=0,
\end{equation}
$$\omega^{2}_{k}(t)=\frac{E(k)}{a^{2}(t)}+m^{2}.\ \  k=0,1,2,\ldots ,$$
and the measure is defined by
$$\begin{array}{rcl}
\int d\mu(\vec k)&=&\sum_{k=0}^{\infty}\sum_{l=0}^{k}\sum_{m=-l}^{l},\  \vec k=(k,l,m),\ E(k)=k(k+2)\ {\rm for}\ \kappa =1,\\
\int d\mu(\vec k)&=&\int_{\mathbb {R}^{3}}{\rm d}^{3}k,\  \vec k\in\mathbb {R}^{3},\ k=|\vec k|,\  E(k)=k^{2}\ {\rm for}\ \kappa =0,\\
\int d\mu(\vec k)&=&\int_{\mathbb {R}^{3}}{\rm d}^{3}k,\  \vec k\in\mathbb {R}^{3},\ k=|\vec k|,\  E(k)=k^{2}+1\ {\rm for}\ \kappa =-1.
\end{array}$$
Each of these Riemannian manifolds have their own generalized eigenfunctions (for details see \cite {jun}).
This system of functions is complete
 and orthonormal, hence we can define a generalized Fourier transform
 $$\begin{array}{rcl}
 \tilde{}: L^{2}(\Sigma) & \to  & L^{2}(\Sigma)\\
                              h & \mapsto & \tilde{h}(\vec k) =\int_{\Sigma}\overline{\phi_{\vec k}(\vec y)}
                              h(\vec y)d^{3}\sigma,
 \end{array}$$
 where $d^{3}\sigma=\sqrt{|s|}dy=(1-\kappa r^{2})^{-\frac{1}{2}}r^{2}\sin\theta drd\theta d\varphi$.
 
 The phase space $(\Gamma, \varsigma)$ of the initial data $\Gamma=C^{\infty}_{0}(\Sigma)\times
 a^{3}(t)C^{\infty}_{0}(\Sigma)$ has the symplectic form
 $$\varsigma(F_{1}, F_{2})=-a^{3}(t)\int_{\Sigma}[q_{1}p_{2}-q_{2}p_{1}]d^{3}\sigma,$$
 for $F_{i}=\left(q_{i} \atop a^{3}(t)p_{i}\right)\in\Gamma$, $i=1,2$.\\
 Following a theorem in \cite{jun} (for the proof see \cite{lr}) we introduce two parameters
 which will serve to describe certain states.
 \begin{veta}
The homogeneous, isotropic quasifree states for the free Klein-Gordon field in a Robertson- Walker
 spacetime are given by the following  2-point function
$$\lambda^{(2)}(F_{1}, F_{2})=\int \langle\overline{\tilde{F}_{1}(\vec k)},
 S(k)\tilde{F}_{2}(\vec k)\rangle ,$$
 $$ S(k)=\left(\begin{array}{cc}
 |p(k)|^{2} & -q{k}\overline {p(k)}\\
 -\overline{q(k)}p(k) & |q(k)|^{2}
 \end{array}\right),$$
 where $p(k)$ and $q(k)$ are (essentially bounded measurable) complex valued functions
 satisfying
 $$\overline{q(k)}p(k)-q(k)\overline{p(k)}=-i.$$
 \end{veta}
 Using these parameters we can according to \cite{jun} write down the formulas defining the adiabatic vacuum states, which in 
 limit $n\to\infty$ gives the Hadamard states.
 \begin{definice}
 An {\em adiabatic vacuum state of order} n is a homogeneous, isotropic Fock state whose 2-point
 function is given by the functions $q(k)=T_{k}(t)$, $p(k)=a^{3}(t)\dot T_{k}(t)$ where $T_{k}(t)$
 is a solution of the differential equation (4) with initial conditions at time t
 \begin{eqnarray}
 T_{k}(t) & = & W^{(n)}_{k}(t)\\
 \dot T_{k}(t) & = & \dot W^{(n)}_{k}(t).
 \end{eqnarray}
 Here,
\begin{equation}
W^{(n)}_{k}(t)=\frac{e^{-i\int_{t_{0}}^{t}\Omega^{(n)}_{k}(t)dt'}}{a^{3/2}(t)
\sqrt{2\Omega^{(n)}_{k}(t)}}
\end{equation}
is iteratively defined by
\begin{eqnarray}
 (\Omega^{[0]}_{k}(t))^{2} & = & \omega_{k}^{2}(t)=\frac{E(k)}{a^{2}(t)}+m^{2} \\
 (\Omega_{k}^{[n+1]})^{2} & = & \omega_{k}^{2}(t)-
 \frac{3}{4}\left(\frac{\dot a(t)}{a(t)}\right)^{2}-
 \frac{3}{2}\frac{\ddot a(t)}{a(t)}
 +\frac{3}{4}\left(\frac{\dot\Omega_{k}^{[n]}(t)}{\Omega_{k}^{[n]}(t)}\right)^{2}
 -\frac{1}{2}\frac{\ddot\Omega_{k}^{[n]}(t)}{\Omega_{k}^{[n]}(t)}.
 \end{eqnarray}
 \end{definice}
 The functions $\Omega_{k}^{[n]}(t)$ are iterative solutions of the equation
 $$ \Omega_{k}^{2}(t) = \omega_{k}^{2}(t)-\frac{3}{4}\left(\frac{\dot a(t)}{a(t)}\right)^{2}
 -\frac{3}{2}\frac{\ddot a(t)}{a(t)}
 +\frac{3}{4}\left(\frac{\dot\Omega_{k}(t)}{\Omega_{k}(t)}\right)^{2}-
 \frac{1}{2}\frac{\ddot\Omega_{k}(t)}{\Omega_{k}(t)},$$
 which is closely linked to the equation (\ref{te}), see \cite{lr}.
 From this definition we see that the adiabatic vacuum states are essentialy dependent on the
 order of the iteration and on the initial time $t$.
 
\section{Main Result}

Now we will say that the {\it Hadamard condition is valid} if
 $$\Omega_{k}^{[n]}(t) {\rm \ exists\ for\ each\ natural\ number\ } n.\ \ \ \ {\rm (H1)}$$
 This means, in particular, that
 $$\Omega_{k}^{[n]}(t) {\rm \ is\ twice\ continously\ differentiable\ \ \ \ \ (H2)}$$
 and
  $$\Omega_{k}^{[n]}(t)>0, \forall t\in\mathbb {R} {\rm \ for\ any\ }n.\ \ \ \ {\rm (H3)}.$$
\begin{veta}
Let $(M,g)$ be a Robertson-Walker spacetime with $a(t)\in C^{2}(\mathbb {R})$. Suppose
that the Hadamard condition is valid. Then $a(t)$ is a smooth function.
\end{veta}
Proof: 
We use the mathematical induction.  Calculating the first iteration we get
$$
(\Omega^{[1]}_{k}(t))^{2} -
\frac{1}{4a^{6}(t)\omega_{k}^{4}(t)}
[4a^{6}(t)\omega_{k}^{6}(t)-$$
$$-3a^{4}(t)\dot a^{2}(t)\omega_{k}^{4}(t)
+6a^{5}(t)\ddot a(t)\omega_{k}^{4}(t)
 +3[E(k)\dot a(t)]^{2}+$$
\begin{equation}
+2E(k)(\ddot a(t)a(t)-\dot a^{2}(t))a^{2}(t)\omega_{k}^{2}(t)
-2E(k)a^{6}(t)\dot a^{2}(t)]=0,
\end{equation}
The expression (10) is linearly dependent on the  highest derivative of $a(t)$ so we can express
$$\ddot a(t)=\frac{1}{2\omega_{k}^{2}(t)(3a^{5}(t)+E(k)a^{3}(t))}
[4a^{6}(t)\omega_{k}^{6}(t)
-3\dot a^{2}(t)a^{4}(t)\omega_{k}^{4}(t)-
3E^{2}(k)\dot a^{2}(t)$$
\begin{equation}
-2E(k)\dot a^{2}(t)a^{2}(t)\omega_{k}^{2}(t)-2E(k)a^{2}(t)\dot a^{6}(t)-
4a^{6}(t)(\Omega^{[1]}_{k})^{2}(t)\omega_{k}^{2}(t)].
\end{equation}
Since $a(t)\in C^{2}(\mathbb {R})$ by hypothesis, the right-hand side has a continous
derivative, hence so has the left-hand side, i. e. $a(t)\in C^{3}(\mathbb {R})$.
This in turn means-by differentiating (11) again-that right hand side is in $C^{2}(\mathbb {R})$
hence $a(t)\in C^{4}(\mathbb {R})$. From (10) we have
$$\left(\Omega^{[1]}_{k}(t)\right)^{2}=\left(2E(k)a^{3}(t)\ddot a(t)\omega_{k}^{4}(t)
-6a^{5}(t)\omega_{k}^{2}(t)\right)\ddot a(t)+$$
$$+({\rm terms\ involving\ only\ }a(t),\dot a(t)),$$
in particular, $\left(\Omega^{[1]}_{k}(t)\right)^{2}$ depends linearly on $\ddot a(t)$.\\
Now assume  $a(t)\in C^{2n}(\mathbb {R})$ and consider the $n$-th iteration ($n\geq 2$)
\begin{equation}
(\Omega_{k}^{[n]})^{2} =\omega_{k}^{2}(t)-
 \frac{3}{4}\left(\frac{\dot a(t)}{a(t)}\right)^{2}-
 \frac{3}{2}\frac{\ddot a(t)}{a(t)}
 +\frac{3}{4}\left(\frac{\dot\Omega_{k}^{[n-1]}(t)}{\Omega_{k}^{[n-1]}(t)}\right)^{2}
 -\frac{1}{2}\frac{\ddot\Omega_{k}^{[n-1]}(t)}{\Omega_{k}^{[n-1]}(t)},
\end{equation}
where
$$\left(\Omega_{k}^{[n-1]}(t)\right)^{2}=F_{n}(a(t),\dot a(t),\ldots, a^{(2n-2)}(t)),$$
and assume that $F_{n}$ depends on $a^{(2n-2)}(t)$ linearly,
$$\left(\Omega_{k}^{[n-1]}(t)\right)^{2}=f_{n}(a(t), \dot a(t),\dots, a^{(2n-3)}(t))a^{(2n-2)}(t)+$$
\begin{equation}
+({\rm terms\ involving\ only\ }a(t), \dot a(t),\dots, a^{(2n-3)}(t)).
\end{equation}
This implies
$$\left[{\left(\Omega_{k}^{[n-1]}(t)\right)^{2}}\right]\ddot\  =f_{n}(a(t), \dot a(t),\dots, a^{(2n-3)}(t))a^{(2n)}(t)+$$
\begin{equation}
+({\rm terms\ involving\ only\ }a(t), \dot a(t),\dots, a^{(2n-1)}(t)),
\end{equation}
with $f_{n}\in C^{\infty}$.\\
By (9)
\begin{equation}
(\Omega_{k}^{[n]})^{2}
-\bigg[\omega_{k}^{2}(t)-
 \frac{3}{4}\left(\frac{\dot a(t)}{a(t)}\right)^{2}-
 \frac{3}{2}\frac{\ddot a(t)}{a(t)}
 +\frac{3}{4}\left(\frac{\dot\Omega_{k}^{[n-1]}(t)}{\Omega_{k}^{[n-1]}(t)}\right)^{2}
 -\frac{1}{2}\frac{\ddot\Omega_{k}^{[n-1]}(t)}{\Omega_{k}^{[n-1]}(t)}\bigg]=0.
 \end{equation}
 Since
 $$\frac{\dot \Omega_{k}^{[n-1]}(t)}{\Omega_{k}^{[n-1]}(t)}-
 \frac{1}{2}\frac{\left[{\left(\Omega_{k}^{[n-1]}(t)\right)^{2}}\right]\ddot\  }{\big(\Omega_{k}^{[n-1]}(t)\big)^{2}},$$
 and
 $$\frac{\ddot \Omega_{k}^{[n-1]}(t)}{\Omega_{k}^{[n-1]}(t)}=
 \frac{1}{2}\frac{\left[{\left(\Omega_{k}^{[n-1]}(t)\right)^{2}}\right]\ddot\  }{\big(\Omega_{k}^{[n-1]}(t)\big)^{2}}
 -\left(\frac{\dot\Omega_{k}^{[n-1]}(t)}{\Omega_{k}^{[n-1]}(t)}\right)^{2},$$
 it follows from (13) and (14) that the left hand side of (15) depends linearly on $a^{(2n)}(t)$,
 $$\left(\Omega_{k}^{[n]}(t)\right)^{2}= ({\rm terms\ involving\ only\ }a(t), \dot a(t),\dots, a^{(2n-3)}(t))-$$
 \begin{equation}
 -\frac{1}{4}\frac{f_{n}(a(t), \dot a(t),\dots, a^{(2n-3)}(t))}{\big(\Omega_{k}^{[n-1]}(t)\big)^{2}}a^{(2n)}(t).
 \end{equation}
 Thus
 $$a^{(2n)}(t)=({\rm terms\ involving\ only\ }a(t), \dot a(t),\dots, a^{(2n-3)}(t), \Omega_{k}^{[n-1]}(t))-$$
\begin{equation}
-4\frac{\left(\Omega_{k}^{[n]}(t)\right)^{2}\left(\Omega_{k}^{[n-1]}(t)\right)^{2}}
 {f_{n}(a(t), \dot a(t),\dots, a^{(2n-3)}(t))}.
 \end{equation}
 The induction hypothesis $a(t)\in C^{2n}(\mathbb{R})$ together with (H2) implies that the right hand
 side of (17) is in $C^{2}(\mathbb {R})$. Thus $a^{(2n)}(t)\in C^{2}(\mathbb {R})$, i. e.
 $a(t)\in C^{2n+2}(\mathbb{R})$.\\
 Besides (17) shows that $\left(\Omega_{k}^{[n]}(t)\right)^{2}$ depends on $a^{(2n)}(t)$ linearly,
 i. e. (13) holds for $n+1$ in the place of $n$. Consequently, by induction on $n$, we conclude that
 $a(t)\in C^{\infty}(\mathbb{R})$.\\
 To make the passage from (16) to (17) completely rigorous, it remains to check that the denominator in (17)
 does not vanish, i. e.
 $$f_{n}(a(t), \dot a(t),\dots, a^{(2n-3)}(t))\not=0.$$
 Observe that, by (11),
 $$f_{2}(a(t), \dot a(t))=\omega_{k}^{2}(t)(3a^{5}(t)-E(k)a^{3}(t)),$$
 while by (16)
 $$f_{n+1} (a(t), \dot a(t),\dots, a^{(2n-1)}(t))=
 -\frac{1}{4}\frac{f_{n} (a(t), \dot a(t),\dots, a^{(2n-3)}(t))}{\big(\Omega^{[n-1]}_{k}(t)\big)^{2}}.$$
 Iteratively the last relation gives ($n\geq 1$)
  $$f_{n+1} (a(t), \dot a(t),\dots, a^{(2n-1)}(t))=
  \left(-\frac{1}{4}\right)^{n-1}f_{2}(a(t),\dot a(t))\prod_{i=1}^{n-1}\frac{1}{\big(\Omega^{[i]}_{k}(t)\big)^{2}}.$$
 The last denominator is nonzero by (H3).\\
 Remark: Observe that it follows from (13) and Theorem 2 that $\Omega^{[n]}_{k}(t)$, are
 in fact, not only $C^{2}$ but $C^{\infty}$.

 \section{Conclusion}
 We have proved that on the Robertson-Walker spacetime the validity of the Hadamard 
 condition for the free scalar quantum fields implies smoothness of the scale factor, i. e.
 together with the paper \cite {lr} we can state that in our case the validity of the Hadamard
 condition is a sufficient condition for the smoothness of the scale factor
 of the Robertson-Walker spacetime.\\
 There are still some open questions, for instance, whether
we can derive the same result for different kinds of quantum fields, e. g.
          Hermite scalar fields, spinor fields, etc., 
or whether it is valid also for others spacetimes than the Robertson-Walker spacetime.

\section{Acknowledgements}
The author acknowledge the support from the M\v{S}MT under project MSM4781305904,
and ESI in Vienna.
Special thanks are to M. Engli\v{s} for helpful discussions.

\end{document}